\documentclass[twocolumn,amsmath,longbibliography,amssymb,superscriptaddress]{revtex4-2}

\usepackage[pdftex]{graphics}
\usepackage{graphicx}
\graphicspath{{figures/}}
\usepackage{hyperref}
\usepackage{xcolor}
\usepackage{physics}
\usepackage{subfig}
\usepackage{bm}
\usepackage{caption}
\usepackage{amsmath} 
\usepackage{mathtools}

\begin{document}
		
\title{Polarization and entanglement spectrum in non-hermitian systems}
\author{Carlos Ortega-Taberner}
\affiliation{Department of Physics, Stockholm University, AlbaNova University Center, SE-106 91 Stockholm, Sweden}
\affiliation{Nordita, KTH Royal Institute of Technology and Stockholm University, SE-106 91 Stockholm, Sweden}

\author{Lukas R\o dland}
\affiliation{Department of Physics, Stockholm University, AlbaNova University Center, SE-106 91 Stockholm, Sweden}

\author{Maria Hermanns}
\affiliation{Department of Physics, Stockholm University, AlbaNova University Center, SE-106 91 Stockholm, Sweden}
\affiliation{Nordita, KTH Royal Institute of Technology and Stockholm University, SE-106 91 Stockholm, Sweden}

\date{\today}
\begin{abstract}
The entanglement spectrum is a useful tool to study topological phases of matter, and contains valuable information about the ground state of the system. Here, we study its properties for free non-Hermitian systems for both point-gapped and line-gapped phases. While the entanglement spectrum only retains part of the topological information in the former case, it is very similar to Hermitian systems in the latter. In particular, it not only mimics the topological edge modes, but also contains all the information about the polarization, even in systems that are not topological.
Furthermore, we show that the Wilson loop is equivalent to the many-body polarization and that it reproduces the phase diagram for the system with open boundaries, despite being computed for a periodic system.
\end{abstract}

\maketitle

\section{Introduction}

Since the seminal work by Kane and Mele~\cite{kane_topological_2005}, non-interacting topological phases of matter have attracted growing interest. 
Such phases are characterized by topological invariants, and are often protected by symmetries. 
An extensive classification of possible phases protected by time-reversal, particle-hole symmetry, or a combination of the two, was done by Schnyder et al in Ref.~\cite{schnyder_classification_2008}, showing that gapped Hamiltonians can only have 10 different symmetry classes. 
However, this classification can be extended by also considering gapless systems and/or lattice symmetries, in particular inversion or rotation symmetries, see for instance Ref.~\cite{chiu_classification_2016} and references therein. 

     Recently non-hermitian topological phases have gained a lot of interest~\cite{bergholtz_exceptional_2021, Ashida_non_2020}. Systems with a non-hermitian Hamiltonian can be used as an effective description for systems with gain or loss. 
    Since the eigenvalues of the Hamiltonian are in general complex, one needs to refine the definition of an energy gap. 
    We distinguish between line-gapped systems, where it is possible to draw a line in the complex energy plane between two bands, and point-gapped systems, where the system has a point gap at $E_0$ if the energy bands don't touch $E_0$.

    Line-gapped and point-gapped systems have to be treated differently when classifying non-hermitian topological phases~\cite{Bernard2002,PhysRevX.8.031079,PhysRevX.9.041015,PhysRevB.99.235112}. 
    The main idea of the classification schemes is to map the Hamiltonian to a Hermitian Hamiltonian, where the classification is known, without changing the topological invariants. 
    The Hermitization process is different for line gap and point gap, which may give different topological invariants for line-gapped systems and point-gapped systems even if they are in the same symmetry class.
    The 10 discrete Altland-Zirnbauer symmetry classes, that are used to classify symmetry-protected topological phases for Hermitian systems, have to be extended to 38 Bernard-LeClair symmetry classes~\cite{Bernard2002, PhysRevX.8.031079, PhysRevX.9.041015}.
    The extension of the symmetry classes implies a change in the topological invariants used to distinguish the different topological phases. 
    For instance, some of the symmetry classes are classified by a $\mathbb Z\oplus \mathbb Z$ invariant, which means that you have two independent topological invariants, see Ref.~\cite{PhysRevX.9.041015}.
    Some of the invariants have natural counterparts in the Hermitian system, while others, like the winding of the complex energy bands, are unique to non-Hermitian systems~\cite{PhysRevX.8.031079,Shen_Topological_2018}.
    The topological invariants that we focus on in this work are the winding number and the polarization. Winding numbers for non-Hermitian systems have been widely used, see for instance references \cite{yin_geometrical_2018,Gong_winding_2018,Jian_winding_2018}. In Hermitian systems one finds different definitions and formulations of the polarization in the literature~\cite{KingSmith_polarization_1993,Vanderbilt_polarization_1993,Resta_geometric_1994,Resta_polarization_1992,resta_quantum_1997,Watanabe_polarization_2018} with different physical interpretations, like the Zak phase \cite{zak_berry_1989}. Their generalizations to non-Hermitian systems have only recently been considered, in particular generalizations of the Zak phase \cite{Wagner_berry_2017,Lieu_topological_2018,Lian_berry_2013}, and very recently of Resta's polarization \cite{Lee_many_2020}.
    Note that an invariant called ``biorthogonal polarization" was also introduced for non-Hermitian systems \cite{Kunst_correspondence_2018,Edvardsson_polarization_2020}. It is, however, not a generalization of the Hermitian polarization, as it is calculated using only the topological zero-energy states. It is different from the polarization studied in this paper, although we will use the same name.
    
    A useful tool to study topological phases of matter is the so-called entanglement spectrum~\cite{li_entanglement_2008}.
It is obtained by  computing the reduced density matrix of a sub-region, and identifying it with the negative logarithm of the so-called `entanglement Hamiltonian'. 
The spectrum of the latter is called entanglement spectrum. 
Although originally introduced in strongly interacting systems, it was soon shown to be useful in non-interacting topological systems as well~\cite{turner_topological_2011,fidkowski_entanglement_2010,yao_2010_entanglement}, where it can be computed very efficiently using the correlation function of the sub-region~\cite{peschel_calculation_2003}.\
In fact, in non-interacting systems one often does not compute the many-body entanglement spectrum at all, but focuses instead on the spectrum of the correlation function of the sub-region, the spectrum of which is in the following called entanglement occupancy spectrum (EOS). 
For Hermitian systems, there is a one-to-one correspondence between the topological edge states of the system with open boundary conditions, and the $1/2$ modes of the EOS. 
Remarkably, the EOS has interesting information about the ground state beyond the mere number of topological edge modes. 
For instance, one can extract the Zak phase from the EOS~\cite{ortega-taberner_relation_2021}. 

It is a non-trivial task to generalize the concept of entanglement spectrum to non-Hermitian systems.
There are several issues, most importantly how to choose the `ground state' for which the reduced density matrix is computed, but also how to properly define the reduced density matrix itself.
This was first studied by Herviou et al~\cite{herviou_defining_2019,herviou_entanglement_2019}, who showed that the EOS correctly reproduced the topology of the periodic boundary condition system for line-gapped phases.
For point-gapped phases, however, their conclusion was that the EOS does not retain information about the topology of the state. 

In this manuscript, we revisit the question of how much information about the `ground state' is encoded in the EOS for non-Hermitian systems. 
We prove that, as in Hermitian systems, the Zak phase is encoded in the EOS for line-gapped phases, though not for point-gapped phases. 
Moreover, we demonstrate that the EOS retains some topological information even in point-gapped phases. 
In particular, it harbors two $\xi=1/2$ modes for each \emph{pair} of topological zero modes in the energy spectrum, i.e. two zero modes localized on opposite edges. 
We also study the different generalizations of the polarization in non-Hermitian systems and investigate their relation. In particular we show how the Wilson loop (or discrete Zak phase) reproduces the topology of the system with open boundaries, even though it is computed for periodic boundary conditions.
This was shown previously for the polarization computed using many-body formalism \cite{Lee_many_2020}.
However, as we show here, it is \emph{not} a many-body effect as previously claimed. 

In section \ref{sec:Intro}, we introduce the model with which we exemplify our results, the SSH chain with added second-neighbor hopping. 
We also review some necessary background of non-hermitian systems and the entanglement spectrum. 
We then proceed to discuss our results first for line-gapped phases in section \ref{sec:LG} and then point-gapped phases in section \ref{sec:PG}. 
Some of the more technical details are moved to the appendices.

\section{Model and background}
\label{sec:Intro}
In this section, we provide the model and some necessary background on non-Hermitian systems and the entanglement occupancy spectrum. 
\subsection{Model}\label{sec:Model}
\begin{figure}[t]
\centering
\includegraphics[width=\columnwidth]{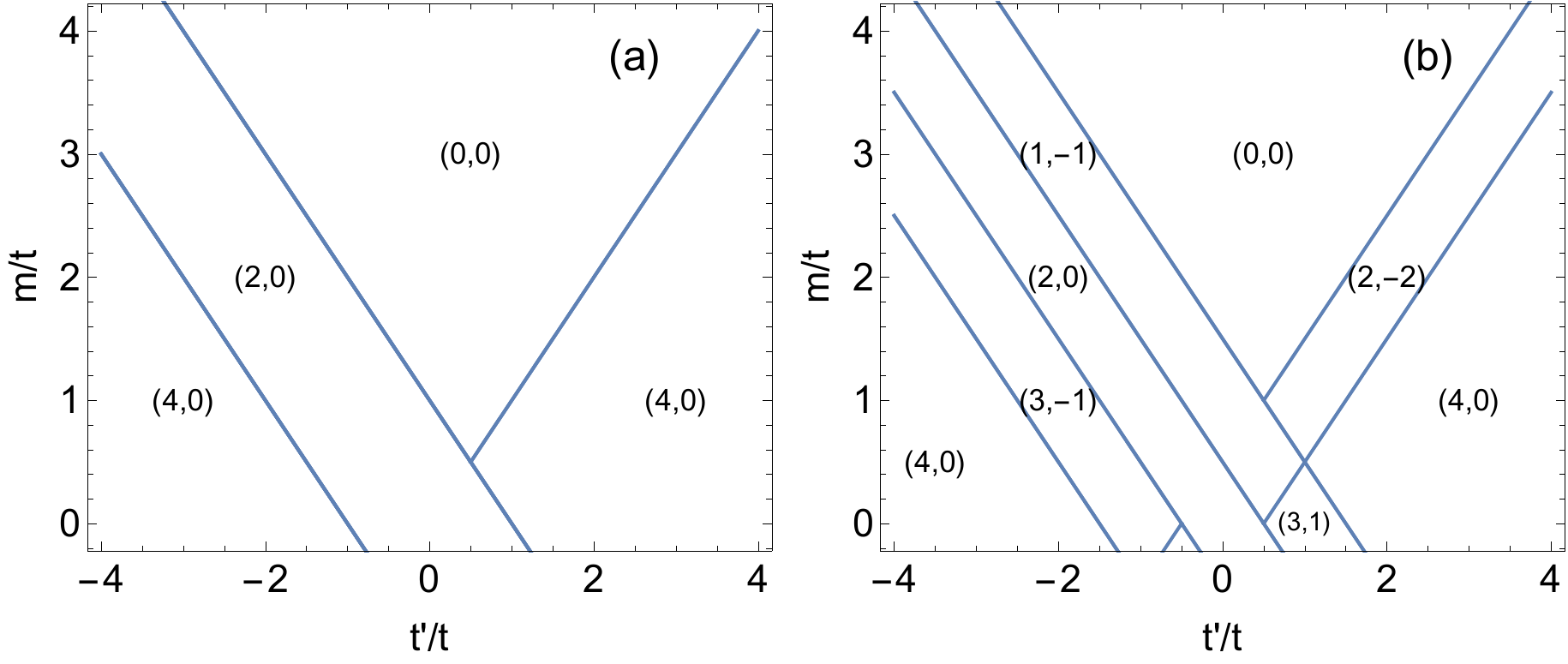}
\caption{Phase diagram for (a) the Hermitian SSH chain and (b) the non-Hermitian model with $\gamma = t/2$. Lines mark the phase-transitions and in parenthesis we show the two winding numbers for each phase, $(\nu,\nu')$. Note that the point-gapped phases are characterized by $\nu \neq 0$.}
\label{fig:phase_diagram}
\end{figure}

One of the most common non-Hermitian models studied is the SSH chain with non-Hermitian hopping. This simple model does not allow us to showcase all our results, so we will extend it by adding second-neighbour hopping. The resulting model is given by
\begin{align}\label{eq:modelH}
H =& \sum_{i\alpha,j\beta} c_{i\alpha}^\dagger H_{ij,\alpha \beta} c_{j\beta},
\end{align}
where 
\begin{align}
H_{ij} =& \frac{1}{2} t \left[(\sigma_x + i \sigma_y)\delta_{i,j+1} + (\sigma_x - i \sigma_y) \delta_{i,j-1} \right] \nonumber\\
&+  \frac{1}{2} t' \left[(\sigma_x + i \sigma_y)\delta_{i,j+2} + (\sigma_x - i \sigma_y) \delta_{i,j-2} \right] \nonumber \\
&+ \mqty( \kappa & m +\gamma \\ m-\gamma & -\kappa)\delta_{ij}.
\label{bdi_model}
\end{align}
For $t' = \gamma = 0$ this is the Rice-Mele model, which reduces to the SSH model for $\kappa=0$. 
With the added second neighbor hopping $t'$, the extended SSH model supports  phases with higher winding numbers
(for a definition of the winding numbers  see the discussion in section \ref{sec:TopInv} below). 
The corresponding phase diagram, with the winding numbers defined as in \eqref{eq:winding2},  is shown in Fig.\ref{fig:phase_diagram}(a). 

There are two different sources of non-hermiticity in our model. 
First, we added the non-Hermitian hopping term $\gamma$, which describes an imbalance in the hopping.  
The on-site potential $\kappa$ provides another non-Hermitian term if chosen to be complex. 
Our model~\eqref{bdi_model} is in the AI symmetry class~\cite{PhysRevX.9.041015} with $\mathcal{T}_+=I$, and it has an additional sub-lattice symmetry, $\mathcal S=\sigma_z$, when $\kappa=0$. 
Since the sub-lattice symmetry commutes with the time-reversal symmetry, we have a \mbox{$\mathbb Z\oplus\mathbb Z$} invariant for the point-gapped phase, and a $\mathbb Z$ invariant for the line-gapped phase when $\kappa=0$~\cite{PhysRevX.8.031079, PhysRevX.9.041015}.
The resulting phase diagram is shown in Fig.\ref{fig:phase_diagram}(b), for a non-hermitian hopping term $\gamma=t/2$ and $\kappa = 0$.

\subsection{Entanglement occupancy spectrum}\label{sec:EOS}
In this subsection, we give a brief review of some basic features of the entanglement spectrum and define what we call the entanglement occupancy spectrum. 
In general, the entanglement spectrum is obtained by dividing the systems into two regions, called in the following A and B, and tracing out the degrees of freedom of part B. 
One can then relate the eigenvalues of the reduced density matrix, $\rho_A$, to the eigenvalues of an `entanglement Hamiltonian' $H_E$ by 
\begin{align}
\label{eq:redDen}
	\rho_A&=e^{-H_E},
\end{align}
and thus obtain the many-body entanglement spectrum. 
Owing to Peschel~\cite{peschel_calculation_2003}, we need in fact not compute the full many-body entanglement spectrum for non-interacting systems, but we can focus on the spectrum of the reduced correlation matrix 
\begin{align}\label{eq:corrMat}
\left(C_A\right)_{(i,\alpha)(j,\beta)}&= \langle c^\dagger_{i,\alpha}c_{j,\beta}\rangle 
\end{align}
where $i$/$j$ denote sites in part $A$, $\alpha$/$\beta$ label internal degrees of freedom, and the expectation value is taken with respect to the ground state. 
To simplify notation, we will suppress the labels for the internal degrees of freedom in the following. 
The entanglement occupancy (EOS) spectrum, given by the set of eigenvalues of the reduced correlation matrix, is sufficient to determine the full entanglement spectrum \eqref{eq:redDen}, see Ref.~\cite{peschel_calculation_2003}. 

Generalizing the entanglement spectrum to non-hermitian systems is not entirely straightforward. 
First, for general non-Hermitian system it is unclear how to choose the appropriate `ground state'~\footnote{Although it is abuse of notation, throughout the paper we call ground state the many-body state chosen to perform the calculations.}. 
The reason why it is difficult to define a ground state is that the energies are complex. One solution is to fill up all the states where the real part of the energy is negative~\cite{Guo_entanglement_2021}, which can lead to several partially filled bands, especially for a real line gap.
The problem with partially filled bands can be solved when considering PT-symmetric models in the PT-unbroken phase~\cite{chen_entanglement_2021,Modak_eigenstate_2021,chang_entanglement_2020} where all the energies are real. 
Recently, there has also been some work arguing for using the steady state solution~\cite{sayyad_entanglement_2021, okuma_quantum_2021} to study the entanglement of non-Hermitian systems rather than using the ground state. This avoids the ambiguity of choosing the ground state, but might not be as useful for studying the band topology.
Since our main goal is to study the band topology in systems without PT symmetry, all the approaches specified above are not suitable.
We, therefore, chose to occupy all the states on one side of the line gap, in line with earlier approaches~\cite{herviou_defining_2019,herviou_entanglement_2019,chen_characterizing_2021,Guo_entanglement_2021}.

The story is more complicated for point-gapped systems, where the two bands merge to one band. Half-filling thus always involves cutting this band in two. There is, of course, not a unique way of doing this, see, for instance, the discussion in Ref.~\cite{Guo_entanglement_2021}. However, we show in Appendix~\ref{app:ground_state} that 
the resulting EOS remains qualitatively the same as long as one chooses the filled states as a connected region.

Another ambiguity involves the generalization of the  density matrix/correlation matrix. 
Note that in non-Hermitian systems the left and right sets of eigenvectors generically do not coincide and we have
\begin{align}
    &H \ket{\psi_\mu^R} = E_\mu \ket{\psi_\mu^R} \nonumber\\
    &H^\dagger \ket{\psi_\mu^L} = E^*_\mu \ket{\psi_\mu^L} 
\end{align}
where we impose the biorthogonalization condition
\begin{equation}
\bra{\psi_\mu^L}\ket{\psi_\nu^R} = \delta_{\mu \nu},
\end{equation}
see Appendix~\ref{app:biorth} for a detailed discussion on how to apply this condition numerically.
This implies that we can define the density/correlation matrix using left eigenstates, right eigenstates or a combination of the two. 
In addition, it is far from obvious that the relation between the entanglement spectrum~\eqref{eq:redDen} and the EOS should still hold for non-Hermitian systems. 
This, however, was proven in Ref.~\cite{herviou_defining_2019}, where they also analyzed the different possible definitions of density/correlation matrix and their respective properties. 
Following their analysis, we focus on the biorthogonal correlation matrix in the remainder of the paper. It is defined as
\begin{align}
C^{LR}_{ij} = \bra{\Psi^L}{c_i^\dagger c_j}\ket{\Psi^R} , 
\end{align}
and is nothing else than the projector onto the occupied bands
\begin{equation}
C^{LR}=  \sum_{n\in occ} \ket{\psi^R_n}\bra{\psi^L_n} . 
\end{equation}
We define the EOS for non-Hermitian systems in a similar way to Hermitian systems, as the set of eigenvalues of the biorthogonal correlation matrix when restricted to a subsystem $A$.
In the following, we drop the $LR$ label and denote the reduced correlation matrix simply by $C_{A}$. 
While we focus solely on the biorthogonal EOS, there are, however, indications that the right-right reduced correlation matrix also harbors information about the system ~\cite{Modak_eigenstate_2021,herviou_entanglement_2019}.

\subsection{Non-hermitian topological invariants}\label{sec:TopInv}

As mentioned above, in section~\ref{sec:Model}, this model for $\kappa=0$ is in a symmetry class which is characterized by a \mbox{$\mathbb Z\oplus\mathbb Z$} invariant for the point-gapped phase. These are two winding numbers, while the line-gapped phases only have one \cite{PhysRevX.8.031079, PhysRevX.9.041015}. 
Consider the Bloch Hamiltonian $H(k) = \boldsymbol{h}(k) \cdot \boldsymbol{\sigma}$. For chiral symmetric systems (which allows us to choose $h_z=$0) we can define two winding numbers~\cite{yin_geometrical_2018} as the winding of two different phases,
\begin{align}
& \tan\phi_1 = \frac{\Re(h_{y})+\Im(h_{x})}{\Re(h_{x})-\Im(h_{y})}\nonumber\\
& \tan\phi_2 = \frac{\Re(h_{y})-\Im(h_{x})}{\Re(h_{x})+\Im(h_{y})},
\end{align}
related to two exceptional points. The winding numbers are then obtained as
\begin{align}\label{eq:winding1}
&\nu_1 = \frac{1}{2\pi}\oint dk \, \partial_k \phi_1 \nonumber\\
&\nu_2 = \frac{1}{2\pi}\oint dk \, \partial_k \phi_2. 
\end{align}
From these the following winding numbers are also defined,
\begin{align}\label{eq:winding2}
&\nu = \frac{1}{2}(\nu_1+\nu_2) \nonumber\\
&\nu' = \frac{1}{2}(\nu_1-\nu_2) ,
\end{align}
where $\nu$ is twice of the winding number usually defined for Hermitian systems and $\nu'=0$ for line gapped systems.

A more physical topological invariant often used in Hermitian systems is the polarization. 
For a system with sub-lattice symmetry, the polarization is a $Z_2$ invariant equivalent to the parity of the winding number. 
There are several different definitions of polarization for a Hermitian system~\cite{KingSmith_polarization_1993,Vanderbilt_polarization_1993,Resta_geometric_1994,Resta_polarization_1992,resta_quantum_1997,Watanabe_polarization_2018}, which lead to objects with slightly different physical meanings. 
The same thing applies to the non-Hermitian case. In the numerical part of this paper we mainly focus on the non-Hermitian biorthogonal generalization of Resta's formula of the polarization~\cite{resta_quantum_1997,Lee_many_2020},
\begin{align}
\label{eq:polarization}
P =& \frac{1}{2\pi} \Im \log \bra{\Psi^L} e^{2\pi i \hat{X}/L}\ket{\Psi^R} \nonumber\\
=& \Im \log \det[S]/2 \pi,
\end{align}
with
\begin{align}
S_{\mu\nu}=\int_0^L dx \, \psi^{L\ast}_{\mu}(x) e^{i2\pi x /L} \psi^R_{\nu}(x),
\end{align}
where $\psi_\mu^{R/L}$ are the single-particle eigenfunctions. For translationally invariant systems this reduces to the finite-size biorthogonal Wilson loop,
\begin{equation}
\label{eq:Wilson}
    W \equiv  \prod_k \bra{u_k^L} \ket{u^R_{k-2\pi/L}}.
\end{equation}
For a two-band model, one finds that
\begin{align}
e^{i2\pi P} = e^{i\frac{L-1}{2}} \prod_k \bra{u^L_k} \ket{u^R_{k-2\pi/L}},
\end{align}
where $\ket{u^R_{k}}$ are the eigenstates of the Bloch Hamiltonian. 
As in the Hermitian case, the polarization only differs from the phase of the Wilson loop by a constant. 
The derivation of both results above can be found in Appendix~\ref{appendix:polarization_zak phase}, together with a more extended discussion on the relation between the different expressions of polarization for non-Hermitian systems.

This equivalence of polarization and Wilson loop applies both to line-gapped and point-gapped systems, although as we will see in section~\ref{sec:PG}, the point-gapped case requires a more careful analysis.

% - % - % - % - % - % - % - % - % - % - % - % - % - % - % - % - % - % - % - % 
\section{Line-gapped systems}
\label{sec:LG}
\begin{figure}
\centering
\includegraphics[width=\columnwidth]{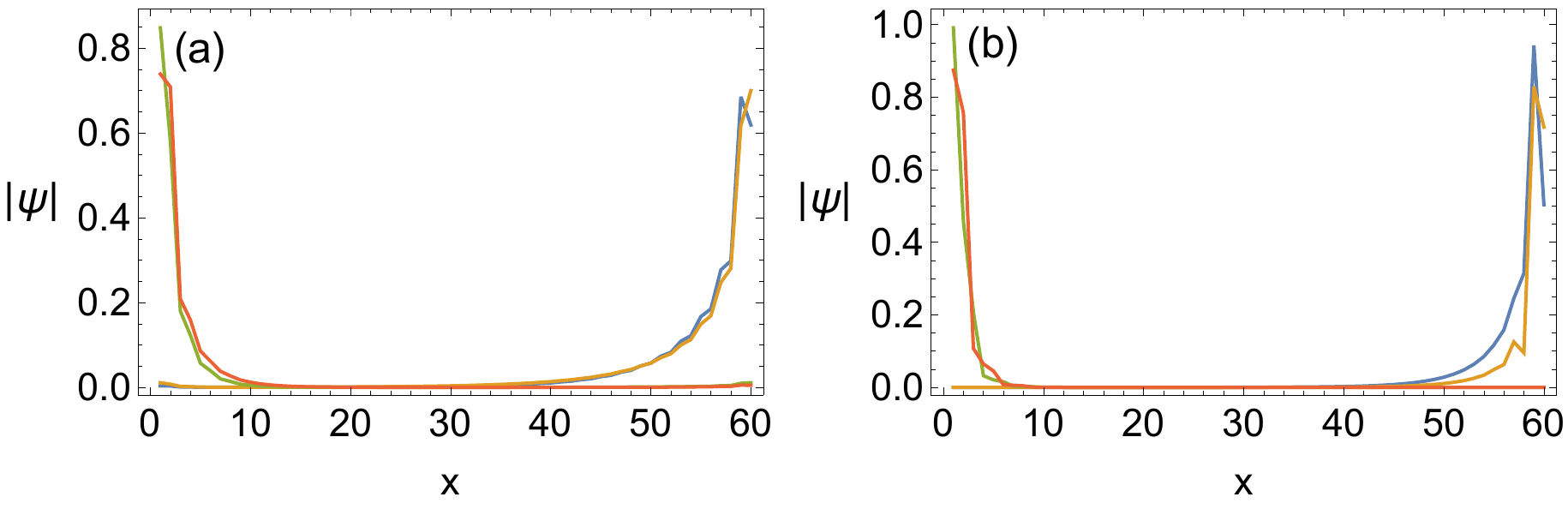}
\caption{We consider the four eigenstates with eigenvalue closest to $1/2$ (midgap) and we plot their norm at each position for parameters $t'=-2t,\kappa=0$, (a) $m=2$ and (b) $m=0$.}
\label{fig:line_gapped_loc}
\end{figure}

We start by looking at line-gapped phases. These are simpler to study  than point-gapped phases as they retain some features of Hermitian phases. 
In this section, we study the EOS and the polarization and show how the relation between them, derived in Ref.~\cite{ortega-taberner_relation_2021} for Hermitian systems, generalizes to the non-Hermitian case for line-gapped phases. More details on the proof can be found in Appendix \ref{appendix:polarization_eos}. In the Hermitian case, we made use of the fact that the eigenstates of $C_A$ are localized on either edge to define 
\begin{equation}\label{eq:chi}
\chi = \sum_{\mu \in L} \xi_\mu ,
\end{equation}
where one sums all eigenvalues of eigenstates that localize in the left boundary of A. 
In the thermodynamic limit, $\chi$ is equal to the polarization. 
In order to generalize $\chi$ to non-Hermitian systems, we must first ensure that the eigenstates of the non-Hermitian $C_A$ are still localized. 
That this is indeed the case is illustrated in~Fig.~\ref{fig:line_gapped_loc},  where we plot a few midgap states (closest to eigenvalue $1/2$) for two sets of parameters corresponding to two different topological cases.

\begin{figure}
\centering
\includegraphics[width=\columnwidth]{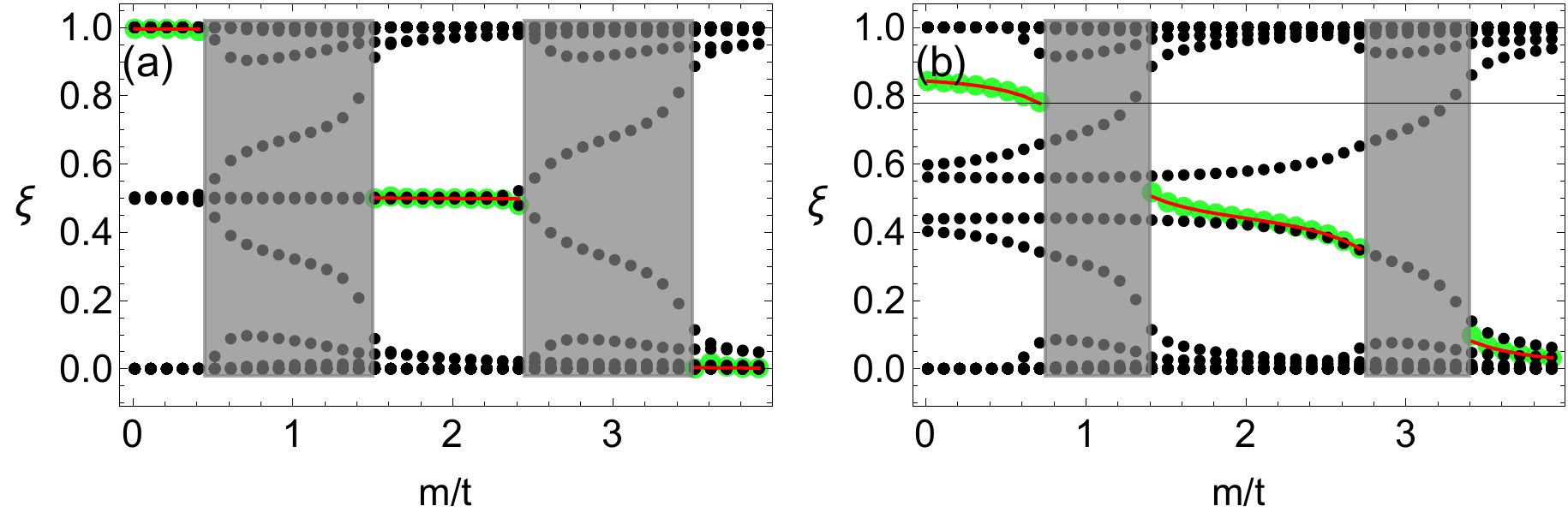}
\caption{In (a) and (b) we show the EOS (black dots), $\chi$ (green dots) and $P$ (red line) for parameters $t'=-2t,\gamma=t/2$ and $\kappa=0,0.3 t$, respectively, for system size $L=120$. The regions with a point gap are greyed out, as they will be discussed in the next section.}
\label{fig:line_gapped_transition}
\end{figure}

In order to demonstrate that $\chi$ reproduces the polarization also in the non-Hermitian case, we look at a cut in the phase diagram in Fig.~\ref{fig:phase_diagram}, which corresponds to taking $t'=-2t$ for the model described above~\eqref{bdi_model}. 
This cut goes through different line-gapped phases which we study in Fig.\ref{fig:line_gapped_transition}. The point-gapped phases are greyed out, because the results of this section only apply to line-gapped phases. 
The line-gapped phases in this plot have an imaginary line gap, thus in order to capture the topology of the bands we choose the ground state by occupying all states with $\Re[E]<0$.

In Fig.\ref{fig:line_gapped_transition}(a) we show the EOS (black dots), the polarization (red line) and $\chi$ (green dots) along this line in the phase diagram ($t'=-2t$). 
Note that we only show the real part of all these objects. Throughout the paper, unless stated otherwise, we only refer to the real part, as the imaginary part is not relevant for most of the results of this paper.
 We see that, as shown before~\cite{herviou_entanglement_2019}, the EOS presents topological $1/2$ modes corresponding to the total winding number $\nu$. Furthermore, we see that the polarization reproduces the parity of the winding number and that indeed, it is equivalent to $\chi$. 
 Choosing $\kappa \neq 0$ breaks sub-lattice symmetry such that there are no topological $1/2$ modes and the polarization is not quantized any longer, see Fig.~\ref{fig:line_gapped_transition}(b). However, the equivalence between the polarization and $\chi$ still holds. In Appendix~\ref{appendix:polarization_eos}, we prove this relation between $\chi$ and the biorthogonal polarization in the non-Hermitian case.
 That the identity between $\chi$ and the polarization should still hold  in non-Hermitian systems is reasonable, but far from obvious. 
The derivation of the proof for Hermitian systems relies on the fact that the Schmidt coefficients are given by the eigenvalues of the reduced density matrix. This relation, however, is broken for non-Hermitian systems~\cite{herviou_entanglement_2019}. 

\begin{figure}
\centering
\includegraphics[width=\columnwidth]{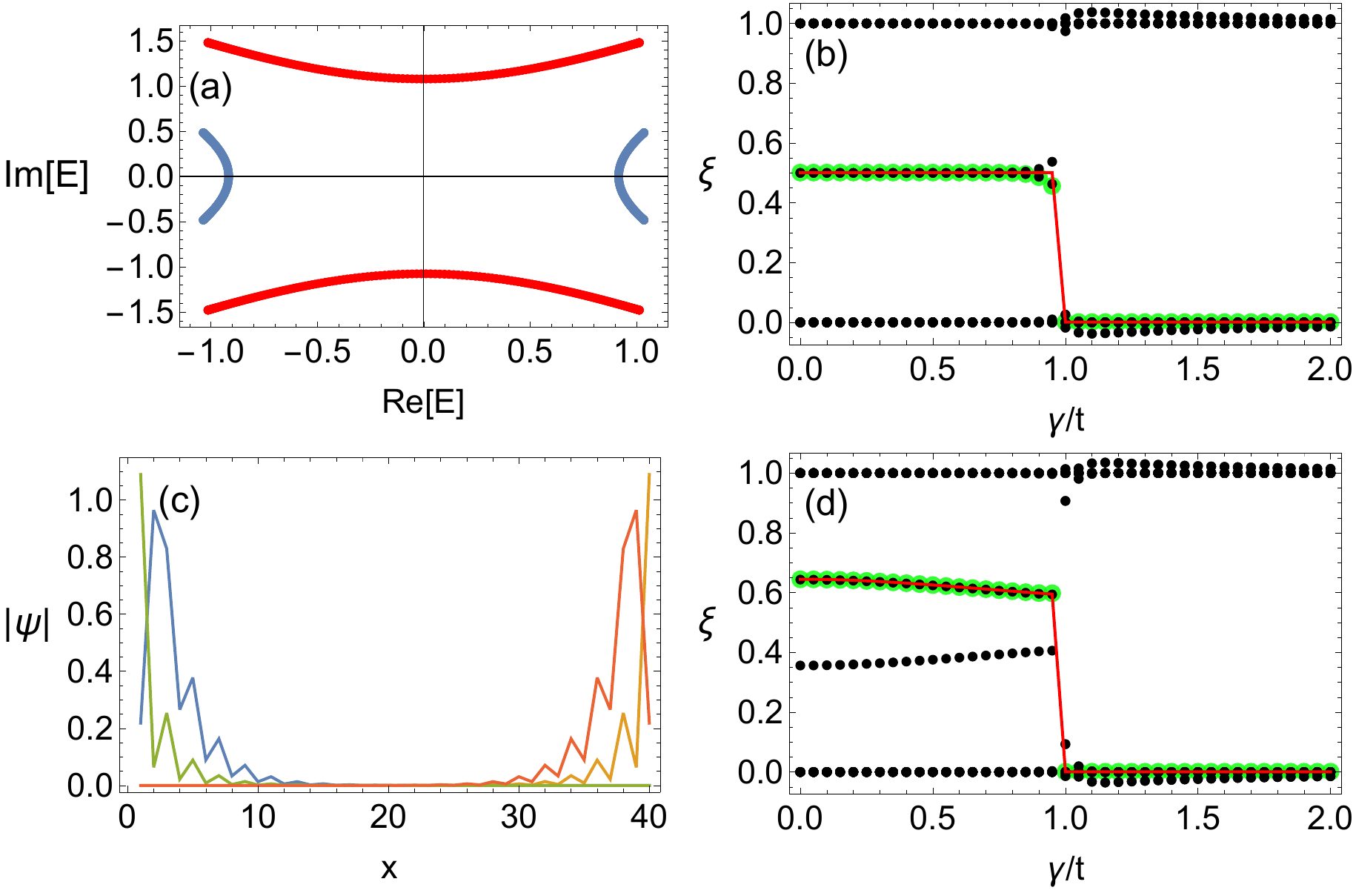}
\caption{(a) Band structures of the system for $\gamma = t/2$ (blue) and $\gamma=3t/2$ (red). (b) Comparison of the polarization (red) and $\chi$ (green) for $m=0,\kappa=0$, the EOS is shown in black (c) localization of the midgap modes for $\gamma=t/2,m=0,\kappa=0$ (d)  Comparison of the polarization (red) and formula (green) for $m=0,\kappa=0.3t$, the EOS is shown in black}
\label{fig:line_gapped_real}
\end{figure}

So far we only looked at line-gapped phases with imaginary line gaps. In Fig.~\ref{fig:line_gapped_real} we show the same results for a transition between an imaginary and real line-gapped phases. 
Choosing $m=t'=0$ in the Hamiltonian \eqref{eq:modelH}, the transition occurs at $\gamma = t$.
The respective band structures are shown in  Fig.\ref{fig:line_gapped_transition}(a), where the real (imaginary) line-gapped phase is plotted in red (blue). 
Since we are interested in studying the topology of the bands, we occupy all states with $\Re[E]<0$ for $\gamma<t$ (imaginary line-gapped) and  $\Im[E]<0$ for $\gamma>t$ (real line-gapped). 
In Fig.\ref{fig:line_gapped_transition} (b) and (d) we show the EOS together with the polarization obtained as Eq.\eqref{eq:polarization} both for the topological phase and the symmetry-broken phase. In both cases, we see a perfect agreement between the polarization (red line) and our formula $\chi$~\eqref{eq:chi} (green dots). 
In Fig.\ref{fig:line_gapped_transition} (c) we show that the midgap modes (those with eigenvalues closes to 1/2 in the EOS) are localized at the respective edges, which is why Eq.~\eqref{eq:chi} can be used in the first place.

% - % - % - % - % - % - % - % - % - % - % - % - % - % - % - % - % - % - % - % 
\section{Point-gapped systems}
\label{sec:PG}
As we saw last section for the line-gapped phases, choosing the ground state of a non-Hermitian system is not as straightforward as in the Hermitian case. 
For point-gapped phases, it becomes even more ambiguous as there are no isolated bands in the system. 
In the following we are going to consider a ground state at half-filling by occupying all states with $\Re[E]<0$,  but the results discussed are independent of how we occupy the states, as long as the filled states form a connected region.

\subsection{Signatures of topology in the EOS}
\label{sec:PGA}
As mentioned before, the EOS was studied for the non-Hermitian SSH model~\cite{herviou_entanglement_2019} and it was shown that for the line-gapped phases it reproduces the topology of the system with open boundary conditions, with a pair of $1/2$ modes present when $\nu=2$, and none where $\nu=0$. Naively one might think that the EOS will have the same number of topological modes as the energy spectrum for open boundary conditions, i.e. a single $1/2$ mode when $\nu=1$ ~\cite{yin_geometrical_2018}, however this is not the case. It was shown that the EOS does not present any topological modes in this point gap phase. 

In fact, looking a bit more into the symmetries of the EOS, one can see that it is impossible for it to harbor a single unpaired topological state. 
To substantiate this claim, we focus on a system with a perfect bipartition, i.e. region A and B have the same size, and translational invariance, which implies $C_B = C_A$. 
Since the full correlation matrix is still a projector, i.e. $C^2=C$,  we find that 
\begin{align}
    C_A C_{AB}= C_{AB}(1-C_A),
\end{align}
where the different matrices are defined by
\begin{align}
&C = \mqty(C_A & C_{AB} \\ C_{BA} & C_A). 
\end{align}
Assuming $\ket{\psi}$ is a topological state of $C_A$, we find 
\begin{align}
C_A (C_{AB} \ket{\psi}) =& C_{AB}(1-C_A)\ket{\psi} \nonumber \\
=& 1/2 C_{AB} \ket{\psi}.
\end{align}
Thus, $C_{AB}\ket{\psi}$ must also be a topological state of $C_A$. 
Assuming locality, the matrix $C_{AB}$ will have its most dominant contributions on the upper right and lower left corners, implying that if $|\psi\rangle$ is localized on one edge of the system, $C_{AB}|\psi\rangle$ is localized on the other. 
Using that topological states are localized on one of the edges, $\ket{\psi}$ and $C_{AB}\ket{\psi}$ cannot represent the same topological state, and thus topological states in the EOS must come in pairs.  

The EOS does not carry any topological signatures for the SSH model~\cite{herviou_entanglement_2019}, but that is no longer the case for systems with higher winding numbers. 
The model we study, with second neighbor hopping $t'=-2t$ and $\gamma=t/2$, harbours two point-gap phases in the regions $m \in [t/2,3t/2]$ and $m\in[5t/2,7t/2]$. 
The second phase is the same topological phase as the point-gapped phase in the SSH model studied in Ref.~\cite{herviou_entanglement_2019}. 

\begin{figure}
\centering
\includegraphics[width=\columnwidth]{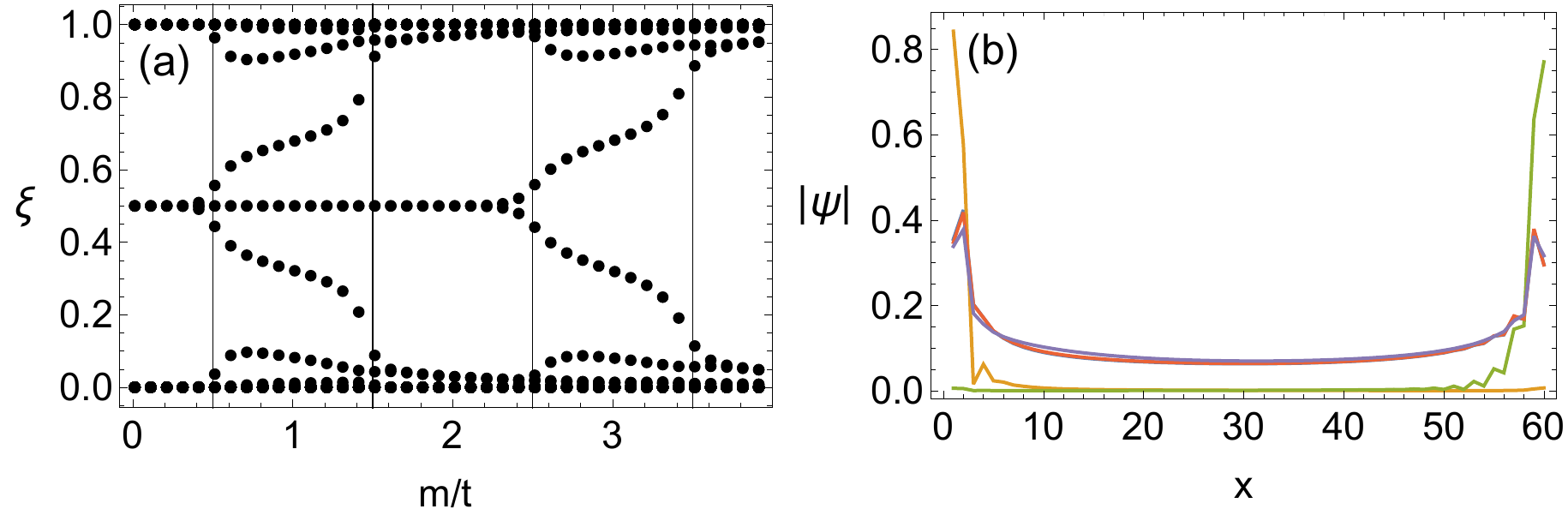}
\caption{(a) EOS for $\gamma = t/2, t' = -2t, \kappa =0$ and $L=120$. Vertical lines mark the phase transitions in the thermodynamic limit. In the point-gapped phase with $\nu=3$ (for $1t/2<m<3t/2$) we observe a pair of topological modes. (b) Norm of the low-lying eigenstates of $C_A$ (closest to $1/2$) in  the same point-gapped phase with $m=t$. Green and orange correspond to the topological modes which fully localize in either edge. Red and Purple are the next two eigenstates, with eigenvalues near the bulk bands at $0$ and $1$. As opposed to the case with a line gap, they have a finite bulk weight. }
\label{fig:point_gapped_transition}
\end{figure}

In Fig.\ref{fig:point_gapped_transition}(a), we show the EOS for the cut in the phase diagram with $\gamma = t/2, t' = -2t$. 
We see that, as discussed previously~\cite{herviou_entanglement_2019} the point-gapped phase of the original SSH (for $5t/2<m<7t/2$) indeed does not host any topological modes. 
Note that, as opposed to the line-gapped phases, the EOS is not degenerate. 
However, the point gapped phase for $t/2<m<3t/2$ does present a pair of topological modes at $1/2$, which are the only degenerate modes.
 Looking at the wavefunction of the midgap states we see that these two states localize in either of the edges. 
 The topological states are the only ones to do so, as all other states in the EOS have weight in both edges, and a non-vanishing weight in the bulk.
 Note that this also prevents us from generalizing Eq.~\eqref{eq:chi} to point-gapped phases. From this result we can argue that the number of topological modes in the EOS is equal to the number of \emph{paired} topological modes in the edge spectrum.

It is noteworthy that the EOS has different degeneracies for the line-gapped and point-gapped phases. 
To better understand this we invoke the same symmetry argument used above. 
In a Hermitian phase the two virtual edges are independent of each other and therefore the eigenstates localize on either edge. 
The operator $\mathcal{S}C_{AB}$, where $\mathcal{S}$ is the sub-lattice symmetry operator, acting on an eigenstate of $C_A$ will result in another eigenstate localize in the opposite edge, leading to a double-degenerate spectrum. 
On the other hand, in the point-gapped phase the two virtual edges are no longer independent of each other. 
The eigenstates become localized on both edges and acting with $\mathcal{S} C_{AB}$ does not necessarily result in a new eigenstate. 
The only double-degenerate states are the topological states, which we did find localize in either edge.

\subsection{Polarization and bulk-boundary correspondence}
\label{sec:PGB}
The polarization, calculated as the many-body average of the position operator \eqref{eq:polarization} has been shown to be an interesting object in non-Hermitian systems ~\cite{Lee_many_2020}. Even though it is computed for periodic boundary conditions, it describes the phase diagram of the open boundary condition system, recovering the bulk-boundary correspondence. In the same article it was also argued that this effect is intrinsically many-body, due to Pauli exclusion principle. 
Even though many-body physics seems to play an important role here, the polarization can still be obtained from the single-particle Hamiltonian, as we show in Appendix~\ref{appendix:polarization_zak phase}. Furthermore, we show there that this polarization is  equivalent to the Wilson loop even in the point-gapped phase. This is remarkable, as the occupied band is discontinuous and one might therefore think that the Wilson loop is not physical. A further discussion on this can be found above Eq.~\eqref{eq:band_disc}.

\begin{figure}
\centering
\includegraphics[width=\columnwidth]{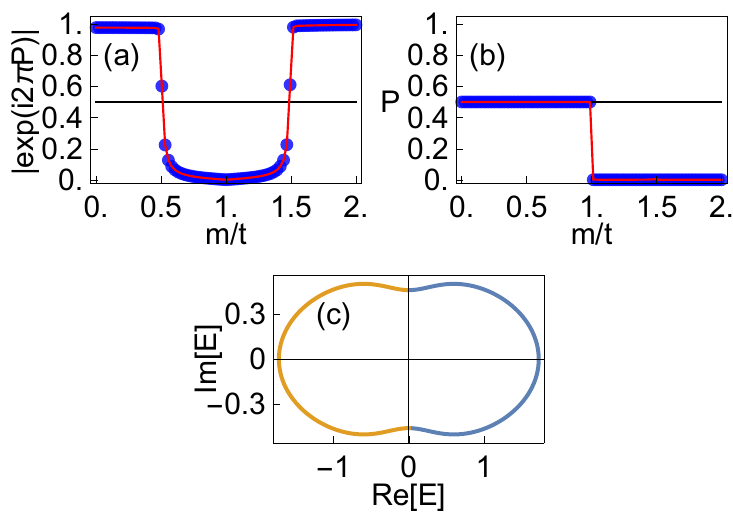}
\caption{For the non-Hermitian SSH model, with $t'=\kappa=0$ and $\gamma=t/2$ and $L=200$. (a) $|\exp[i 2\pi P]|$ and (b) real part of $P$. In red we obtain the polarization using Resta's formula and in blue using the Wilson loop. (c) Energy band in the complex plane for the same parameters in the point-gapped phase with $m=0.8t$. In yellow we show the occupied states used to compute the polarization.}
\label{fig:point_gapped_polarization}
\end{figure}

For the following discussion, we focus on a cut through the phase diagram of the non-Hermitian SSH chain, with $t'=\kappa=0$ and $\gamma=t/2$.
First we show $|\exp[i 2\pi P]|$ in Fig.~\ref{fig:point_gapped_polarization}(a). In the Hermitian case, $P$ is purely real and therefore the absolute value should be $1$, with any deviations coming from finite-size effects. This is still true for the line-gapped phases of the non-Hermitian SSH. In the point-gapped phase, however, this absolute value tends to $0$ in the thermodynamic limit. We plot both the result from Resta's formula \eqref{eq:polarization} (red line) and from the Wilson loop \eqref{eq:Wilson} (blue dots), which are identical. 

In order to understand why the exponential vanishes, we have a closer look at the Wilson loop. In the point-gapped phase there is a discontinuity of the occupied band in momentum space, as it is shown in Fig.\ref{fig:point_gapped_polarization}(c). At this discontinuity the two bands switch and at this point we have
\begin{equation}
\label{eq:band_disc}
\ket{u^R_{occ,k-2\pi/L}} \approx \ket{u^R_{emp,k}} -\frac{2\pi}{L} \partial_k \ket{u_{emp, k}^R}. 
\end{equation}
Therefore
\begin{multline}
   \bra{u_{occ, k}^L} \ket{u_{occ, k-2\pi /L}^R} \approx \\ \bra{u_{occ, k}^L} \ket{u_{emp, k}^R} -\frac{2\pi}{L}  \bra{u_{occ, k}^L} \partial_k \ket{u_{emp, k}^R}. 
\end{multline}
The first term vanishes and therefore the Wilson loop has a constant term $2\pi/L$, while all other terms are close to $1$. In line-gapped phases, on the other hand, there is no such discontinuity and all terms in the Wilson loop are close to $1$. Thus, the imaginary part of the polarization distinguishes between line and point-gapped phases. 

The real part of the polarization, shown in Fig.\ref{fig:point_gapped_polarization}(b) is substantially different, as it captures the open boundary condition phase diagram. We show the result from Resta's formula \eqref{eq:polarization}
as well as the result from the Wilson loop, using the single-particle formulation. These two methods give the same result as the one obtained previously using a many-body formulation \cite{Lee_many_2020}.

\subsection{Extended Hermitian Hamiltonian}
In section \ref{sec:PGA}, we showed that the EOS displays $\min(N_L,N_R)$ pairs $1/2$ modes, where $N_{L/R}$ are the numbers of zero modes at the left/right edge. One may wonder if it is also possible to gain information about the asymmetry, i.e. $|N_L-N_R|$ from the EOS. Below, we show that this indeed can be done, but we need to extend the Hamiltonian to obtain an effective Hermitian system. For the latter, the corresponding EOS will have $N_L+N_R$ pairs of $1/2$ modes. Comparing with the original model allows us to extract the number of edge modes (though not their localization) that are only present on one of the edges.

According to Ref.s~\cite{PhysRevX.8.031079, PhysRevX.9.041015}, a non-Hermitian Hamiltonian H with a point gap will have the same classification as the Hermitian Hamiltonian given by
\begin{equation}
\tilde{H} = \mqty(0 & H \\ H^\dagger & 0).
\end{equation}
Consider a Hamiltonian with open boundary conditions, such that we have topological zero-energy edge states $\ket{\psi_\mu^{{L/R}}}$. The number of topological states of the Hermitian Hamiltonian is then doubled as
\begin{align}
 &\mqty(0 & H \\ H^\dagger & 0)\mqty(\ket{\psi_\mu^L} \\ 0) = 0 \nonumber \\
&  \mqty(0 & H \\ H^\dagger & 0)\mqty( 0\\\ket{\psi_\mu^R} ) = 0 .
\end{align}
Thus, the associated winding number gives the total number of zero-modes of the non-Hermitian Hamiltonian.

The entanglement spectrum of the new Hamiltonian is also related to the winding number, since it is Hermitian. There is then a qualitative difference between the entanglement spectrum of the original and the extended ground states. While for the line-gapped phases the number of topological states is simply doubled from those of the original system, for the point-gapped phase new topological states appear in the extended system.

\begin{figure}[tbh]
\centering
\includegraphics[width=\columnwidth]{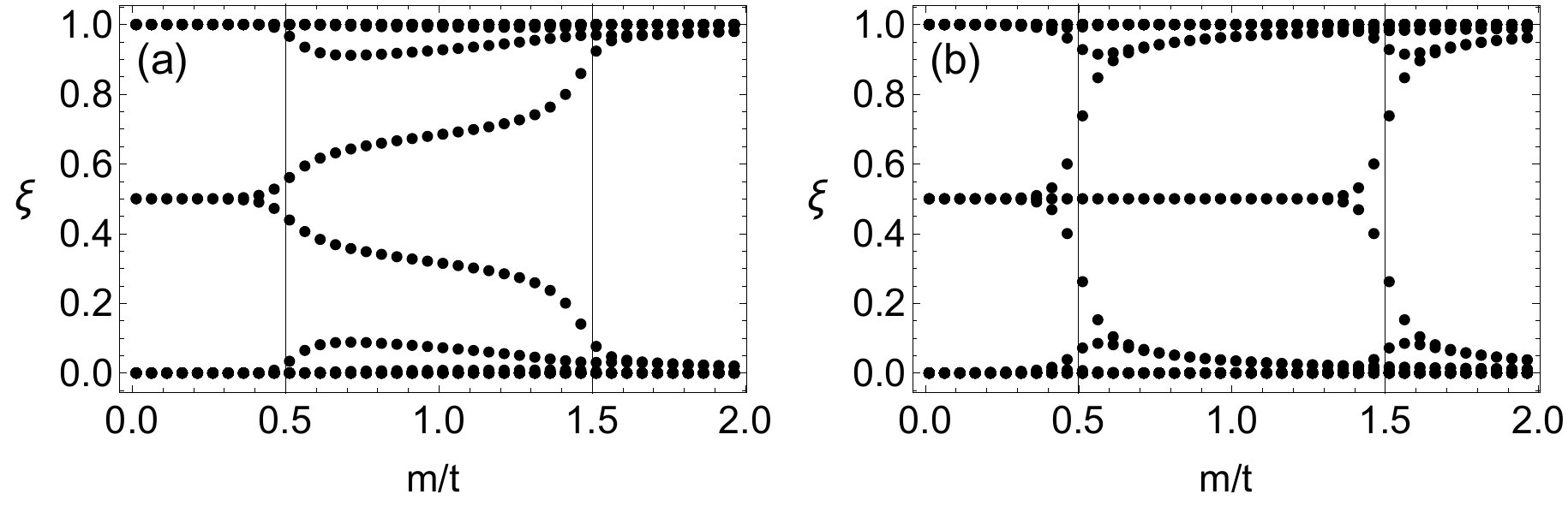}
\caption{EOS for the non-Hermitian SSH, $t'=\kappa=0$ and $\gamma=t/2$. (a) the original non-Hermitian system and (b) for the extended Hermitian system.}
\label{fig:extended_system}
\end{figure}

We showcase this in Fig.~\ref{fig:extended_system}, for the non-hermitian SSH chain, setting $t'=\kappa=\gamma=0$ and varying $m$ in our model Hamiltonian~\eqref{eq:modelH}.
Note that this is the same system as studied in \cite{herviou_entanglement_2019}. In (a) we plot the EOS of the original system,  displaying two $\xi=1/2$ modes in the topological phase ($m<t/2$), which are split in the point-gapped phase$(t/2<m<3t/2)$, and finally merge with the bulk modes at 0 and 1 in the trivial phase ($m>3t/2$).
In Fig.~\ref{fig:extended_system}(b), we plot the EOS of the extended system ($t'=-2$), where we instead find 4/2/0 $\xi=1/2$ modes respectively for the topological/point-gapped/trivial phase. 
% - % - % - % - % - % - % - % - % - % - % - % - % - % - % - % - % - % - % - % 
\section{Conclusion}
Recently there has been growing interest in using the entanglement spectrum and polarization in the study of topology in non-Hermitian systems. Regarding the entanglement spectrum, or more correctly the EOS, it was shown that the line-gapped phases reproduced the topology of the energy spectrum, while the EOS of point-gapped phases was thought to be featureless \cite{herviou_entanglement_2019}. We showed that, although it does not have the full information about the topology of the system, the EOS of point-gapped phases does retain some information about the topology. Regarding the polarization, it was recently shown that a biorthogonal generalization of Resta's polarization, obtained from a many-body numerical calculation, reproduced the topology of the system with open boundaries, recovering the bulk-boundary correspondence \cite{Lee_many_2020}. Here we showed how this polarization can be obtained (more easily) using single-particle physics, and how it reduces to the phase of the Wilson loop for translationally invariant systems. Furthermore we showed that the polarization is encoded in the entanglement spectrum for line-gapped phases.

\emph{Outlook:} We showed that the entanglement spectrum of point-gapped phases encodes more information than was previously thought, but further work is warranted in this direction. In particular, one would like to find a more direct relation between the topological modes in the entanglement spectrum and the ones in the energy spectrum, which does not rely on a Hermitian extension. It is also unclear how to link the topological invariants to certain features of the EOS. One might speculate that combining the information of $C^{LR}$ with those of $C^{RR}$ or $C^{LL}$ could prove useful. 

In this paper we only consider systems with periodic boundary conditions, and it is unclear what information is encoded in the EOS for open boundary conditions.
Furthermore, recently it has been shown that the many-body polarization obtained for open boundary conditions has information about the system with periodic boundary conditions \cite{alsallom_skin_2021}.
In this paper we simplified the calculation of the many-body polarization to single-body physics, the same simplification might be feasible for open-boundary conditions. 

% - % - % - % - % - % - % - % - % - % - % - % - % - % - % - % - % - % - % - % 

\acknowledgments
{\em Acknowledgments.-- }
The research in this grant was supported by the Swedish Research Council under grant no. 2017-05162 and the Knut and Alice Wallenberg foundation under grant no. 2017.0157.

\bibliographystyle{unsrt}
\bibliography{references.bib}

\appendix

\section{Properties of Non-Hermitian systems}

As opposed to the Hermitian case, non-Hermitian matrices have left and right-eigenvectors that can be different. 
In this appendix we introduce notation of the right and left eigenvectors and discuss how the biorthogonalization affects the diagonal fermionic operators. 
We consider a free Hamiltonian given by
\begin{align}
H =& \sum_{ij} c_i^\dagger h_{ij} c_j \nonumber \\
=& \phi^\dagger h \phi.
\end{align}
The single-particle Hamiltonian is non-hermitian and admits two sets of eigenvectors, which we call left and right eigenvectors
\begin{align}
&hR = RE \nonumber\\
&Lh = EL.
\end{align}
$E$ is a diagonal matrix and the biorthogonality condition imposes that $LR=I$. The numerical method employed to impose biorthogonalization is described in Appendix \ref{app:biorth}. The free Hamiltonian is then diagonalized by the transformation
\begin{align}
&\phi^\dagger L^{-1} = \psi_L^\dagger \nonumber\\
&R^{-1}\phi  = \psi_R, 
\end{align}
such that
\begin{align}
H =& \psi^\dagger_L E \psi_R \nonumber\\
=& \sum_\mu \bar{\gamma}_{\mu} E_{\mu\mu} \gamma_\mu.
\end{align}

The new operators fulfill the conditions
\begin{align}
\{ \bar{\gamma}_\mu, \gamma_\nu \} =& \{ \sum_i c_i^\dagger [L^{-1}]_{i\mu} , \sum_{j} [R^{-1}]_{\nu j} c_j \}\nonumber \\
=& [R^{-1}L^{-1}]_{\nu\mu}\nonumber \\
=& \delta_{\mu\nu},
\end{align}
as well as $\{\bar{\gamma}_\mu,\bar{\gamma}_\nu\}=\{ \gamma_\mu , \gamma_\nu\}=0$, so even though the two operators are not conjugate of each other they behave as usual fermionic operators. 
\section{Biorthogonal diagonalization of non-Hermitian matrices}\label{app:biorth}

Since the right and left eigenvectors are obtained numerically as two different sets of states, common numerical methods do not produce eigenstates that fulfill the biorthogonal condition by default. 
In this appendix, we described how it can be imposed by transforming the eigenvectors.
A non-Hermitian matrix, $H$, has two eigenvalue equations
\begin{align}
& HR = RE \nonumber\\
& LH = LE,
\end{align}
where $L,R$ are the collection of left and right eigenvectors and $E$ is a diagonal matrix with the eigenvalues of $H$. To have a biorthogonal diagonalization we have to enforce $LR=I$, however in general $LR=Y$. If there are no degeneracies $Y$ is a diagonal matrix and to enforce biorthonormality we just have to normalize each eigenvector. When degeneracies are present $Y$ is a block-diagonal matrix, with each block corresponding to the degenerate space. In order to normalize the eigenstates we perform a LU decomposition $Y=Y_L Y_U$. We now define the biorthogonal set of eigenvectors as
\begin{align}
& L = Y_L L' \nonumber\\
& R = R' Y_U, 
\end{align}
which fulfill $L'R'=I$. Note that we have to solve the equations above to find the transformed eigenvectors, as numerically inverting the matrices $Y_L$ and $Y_R$ can be problematic. The eigenvalue equation for the right eigenvectors now becomes
\begin{align}
HR'Y_U = R' Y_U E.
\end{align}
At first glance it does not seem that $R'$ are eigenvectors of $H$. However, there is a relation between $Y_U$ and $E$. In the space with no degeneracies $Y_U$ is diagonal, and for the degenerate blocks of $Y_U$, $E$ is an identity with a constant factor. Therefore we have that $[Y_U,E]=0$, and $R'$ fulfills the right eigenvalue equation. The same argument goes for $L'$.

\section{Non-Hermitian Zak phase and Polarization}
\label{appendix:polarization_zak phase}

In Hermitian systems one finds in the literature different definitions of the polarization which are related to one another~\cite{KingSmith_polarization_1993,Vanderbilt_polarization_1993,Resta_geometric_1994,Resta_polarization_1992,resta_quantum_1997,Watanabe_polarization_2018}. In this appendix we consider the generalization of the different definitions of the polarization to the non-Hermitian case, and study the relations among them. In particular we show that the Zak phase~\eqref{eq:zak} and Resta's polarization~\eqref{eq:resta} are equal to the phase of the Wilson loop~\eqref{eq:wilson}, and the polarization obtained by flux insertion~\eqref{eq:pol_flux}

First we consider the biorthogonal generalization of the Zak phase, defined as
\begin{align}
\label{eq:zak}
\gamma = i\int_0^{2\pi} dk \bra{u^L_k} \partial_k \ket{u^R_k}.
\end{align}
In this continuous form it has been used previously for non-Hermitian systems~\cite{Wagner_berry_2017,Lieu_topological_2018,Lian_berry_2013}. 
In the same way as for Hermitian systems, this continuous formulation can be discretized to
\begin{align}
e^{i\gamma} =& \prod_k e^{-\bra{u^L_k}  \ket{u^R_k}}e^{ \bra{u^L_k}  \ket{u^R_{k-2\pi/L}}} \nonumber\\
=& \prod_k e^{\bra{u^L_k}  \ket{u^R_{k-2\pi/L}}-1} \nonumber\\
=& \prod_k \bra{u^L_k}  \ket{u^R_{k-2\pi/L}},
\end{align}
where we define the last expression as the Wilson loop
\begin{equation}
\label{eq:wilson}
    W \equiv \prod_k \bra{u^L_k}  \ket{u^R_{k-2\pi/L}},
\end{equation}
in order to distinguish from equation \eqref{eq:zak}.
In the point gapped phases studied in this paper the occupied band is discontinuous in momentum space and therefore Eq.\eqref{eq:zak} is not well defined. However the Wilson loop can still be applied, leading to interesting features that we discuss in section \ref{sec:PGB}.

Another expression for the polarization found in the literature is the one due to Resta \cite{resta_quantum_1997}. In its biorthogonal form it is defined as
\begin{align}
\label{eq:resta}
P = \frac{1}{2\pi} \Im \log \bra{\Psi^L} e^{2\pi i \hat{X}/L}\ket{\Psi^R}.
\end{align}
which has already been employed for non-Hermitian systems \cite{Lee_many_2020}, although it was only obtained via a many-body calculation. To establish its equivalence with the Zak phase (or Wilson loop) one can use similar arguments as for the Hermitian case. 

Consider the many-body ground state given by occupying the Bloch states
\begin{align}
\bra{X}\ket{\Psi^R} = A \prod_{k\mu} \psi_{k\mu R}(x_{k \mu}).
\end{align}
We can define the state $\ket{\tilde{\Psi}^R}  = e^{2\pi i \hat{X}/L}\ket{\Psi^R}$, or equivalently, redefine the bloch states as $\tilde{\psi}_{k\mu R}(x)  = e^{2\pi i x/L}\psi_{k\mu R}(x)$. The polarization is now a simple state overlap
\begin{align}
e^{i2\pi P} = \bra{\Psi^L} \ket{\tilde{\Psi}^R}. 
\end{align}
Now we can use the same trick as in the Hermitian case, that this overlap is equal to the determinant of the overlap matrix \cite{Plasser_efficient_2018}, to obtain,
\begin{align}
e^{i2\pi P} = \det[S], 
\end{align}
where
\begin{align}
S_{k\mu,p\nu}=\int_0^L dx \, \psi^*_{k\mu L}(x) e^{i2\pi x /L} \psi_{p\nu R}(x)
\end{align}
runs through the indeces of the occupied states. 

Just as in the Hermitian case only a few matrix elements are non-zero, only those with $p=k-2\pi/L$, due to the biorthogonality of the Bloch states. Using this we can factorize the determinant into
\begin{align}
\det[S] = e^{i\frac{L-1}{2}M}\prod_k \det[S_{k,k-2\pi/L} ,]
\end{align}
where $M$ is the number of occupied bands and the determinant is now performed only on the orbital indeces. Note that the constant term is missing in the original paper~\cite{resta_quantum_1997}. It is just a minus sign from the Levi-Civita symbol that comes from expanding the determinant. With only one occupied band we have 
\begin{align}
e^{i2\pi P} = e^{i\frac{L-1}{2}} \prod_k \bra{u^L_k} \ket{u^R_{k-2\pi/L}},
\end{align}
which, other than the constant factor, is indeed the Wilson loop, which in the thermodynamic limit will be also equivalent to the Zak phase. Note that this equivalent between the Resta polarization and the Wilson loop is also valid for point-gapped systems. 

Finally, we now consider the polarization obtained when flux is inserted via twisted boundary conditions \cite{Watanabe_polarization_2018,Zelatel_flux_2014}. In the following we follow the nomenclature introduced by reference \cite{Watanabe_polarization_2018}. For non-Hermitian systems we define
\begin{align}
\label{eq:pol_flux}
\tilde{P} = \int_0^{2\pi} \frac{d\phi}{2\pi} \bra{\tilde{\Phi}^L (\theta)} \partial_\theta \ket{\tilde{\Phi}^R (\theta)},
\end{align}
where $\ket{\Phi^{L/R}(\theta)}$ is the many-body ground state and $\theta$ the flux inserted. Note that, as in the case of the Zak phase, the derivative is not well defined for point-gapped phases, but the discrete formula can still be applied.

Consider a system with 1-site translational invariance,
\begin{align}
\tilde{H}_\theta = \sum_{ij} c^\dagger_{i \alpha} h_{ij,\alpha \beta}(\theta) c_{j\beta},
\end{align}
where latin indices indicate position and greek indices indicate orbital. The Hamiltonian is diagonalized by,
\begin{align}
\gamma_{k^\theta \mu} = \frac{1}{\sqrt{L}} \sum c_{j \alpha} e^{-i k^\theta j} [R^{-1}]_{ \mu\alpha} \nonumber\\
\bar{\gamma}_{k^\theta \mu} = \frac{1}{\sqrt{L}} \sum c^\dagger_{j \alpha} e^{i k^\theta j} [L^{-1}]_{ \alpha\mu} 
\end{align}
where $k^\theta = k + \theta/L$.
The many-body ground state with flux inserted is analogous to the one obtained for the Hermitian system in reference \cite{Watanabe_polarization_2018},
\begin{align}
\ket{\tilde{\Phi} (\theta)} = e^{-i(L-1)M\theta/2} \prod_{k^\theta}\prod_{\mu=1}^M \bar{\gamma}_{k^\theta \mu}\ket{0},
\end{align}
where $M$ is the number of occupied bands and the prefactor is needed to have $\ket{\tilde{\Phi} (\theta+2\pi)} = \ket{\tilde{\Phi} (\theta)}$. Before considering the general many-body ground state consider the Berry connection of the simple state $\bar{\gamma_2}\bar{\gamma}_1 \ket{0}$,
\begin{align}
\bra{0} \gamma_1 \gamma_2 \partial_\theta (\bar{\gamma_2}\bar{\gamma_1}) \ket{0} =& \bra{0} \gamma_1 \gamma_2( \partial_\theta \bar{\gamma_2})\bar{\gamma_1} \ket{0}  \nonumber\\
&+ \bra{0} \gamma_1 \gamma_2 \bar{\gamma_2}( \partial_\theta\bar{\gamma_1}) \ket{0} \nonumber\\
=&\bra{0} \gamma_1 \bar{\gamma_1}  \gamma_2( \partial_\theta \bar{\gamma_2})\ket{0}  \nonumber\\
&+ \bra{0} \gamma_1 \gamma_2 \bar{\gamma_2}( \partial_\theta\bar{\gamma_1}) \ket{0} \nonumber\\
=&\bra{0}  \gamma_2( \partial_\theta \bar{\gamma_2})\ket{0}  \nonumber\\
&+ \bra{0} \gamma_1 ( \partial_\theta\bar{\gamma_1}) \ket{0} 
\end{align}
For a general state we can continue this procedure and reach
\begin{align}
\tilde{P} =& \int_0^{2\pi} \frac{d\theta}{2\pi} i \bra{\tilde{\Phi}^L (\theta)} \partial_\theta \ket{\tilde{\Phi}^R (\theta)} \nonumber\\
=& \frac{L-1}{2}M + i  \int_0^{2\pi} \frac{d\theta}{2\pi} \sum_{k\mu} \bra{0} \gamma_{k^\theta \mu} \partial_\theta \bar{\gamma}_{k^\theta \mu}\ket{0} \nonumber\\
=& \frac{L-1}{2}M + i  \int_0^{2\pi} \frac{d\theta}{2\pi} \sum_{k\mu} \bra{\psi _{k^\theta\mu}^L} \partial_\theta \ket{\psi_{k^\theta \mu}^R}\nonumber \\
=& \frac{L-1}{2}M + i  \int_0^{2\pi} \frac{d\theta}{2\pi} \sum_{k\mu} \bra{k^\theta}\partial_\theta \ket{k^\theta}
\bra{u _{k^\theta\mu}^L} \ket{u_{k^\theta \mu}^R}\nonumber \\
&+ i  \int_0^{2\pi} \frac{d\theta}{2\pi} \sum_{k\mu} \bra{k^\theta} \ket{k^\theta}
\bra{u _{k^\theta\mu}^L}\partial_\theta \ket{u_{k^\theta \mu}^R} ,
\end{align}
where $\ket{\psi_{k \mu}^R}=\ket{k}\otimes \ket{u_{k \mu}^R}$ are the Bloch states. Using that
\begin{align}
\bra{k^\theta}\partial_\theta \ket{k^\theta}=& \frac{1}{L}\sum_{x=0}^{L-1} e^{-ik^\theta x}(ix/L)e^{ik^\theta x} \nonumber\\
=& i\frac{L-1}{2L}
\end{align}
the second term becomes $-\frac{L-1}{2}M$ and cancels with the first one. For the third term, with a change of variable we obtain
\begin{align}
\tilde{P} =  i  \int_0^{2\pi} \frac{dk}{2\pi} \sum_{\mu} 
\bra{u _{k\mu}^L}\partial_k \ket{u_{k \mu}^R} .
\end{align}
This polarization is indeed equal to the Zak phase, and Wilson loop in the discretized case.

\section{Relation between the EOS and the biorthogonal polarization for line-gapped systems}
\label{appendix:polarization_eos}

The relation between the EOS and the polarization described for hermitian systems in previous work can be generalized to non-Hermitian systems in the line gapped case. 
Let us first consider the many-body ground state. 
For non-Hermitian systems, the left and right many-body ground state are distinct, and can be decomposed using a Schmidt decomposition
\begin{align}
&\ket{\Phi^R} = \sum_\mu s_\mu  \ket{\psi_\mu}_A \otimes \ket{\psi_\mu}_B. \nonumber\\
&\ket{\Phi^L} = \sum_\mu l_\mu  \ket{\phi_\mu}_A \otimes \ket{\phi_\mu}_B,
\end{align}
where all sets of $\ket{\phi/\psi_\mu}_{A/B}$ are orthogonal and span the subspace $A/B$. The main difference with the Hermitian case is that now the left and right ground states can admit two different Schmidt decompositions, in particular $\ket{\phi_\mu}_{A/B}$ and $\ket{\psi_\mu}_{A/B}$ are different basis sets. We now insert a flux via twisted boundary conditions \cite{Zelatel_flux_2014,Watanabe_polarization_2018}. In the thermodynamic limit we can assume that the fluctuations at the two virtual cuts are independent from each other and therefore the Schmidt states can be characterized by two quantum numbers associated to each virtual cut, $\ket{\psi_\mu}_A = \ket{p_\mu,q_\mu}_A$ \cite{Zelatel_flux_2014}, where $p_\mu,q_\mu$ are the quantum numbers that characterize the state near each virtual cut. These quantum numbers are not used explicitly in the following, but this decomposition is essential to argue that one can define the number of electrons near the seam
\begin{equation}
\hat{N}^{A_L}\ket{\psi_\mu}_A = N^{A_L}_{\psi_\mu} \ket{\psi_\mu}_A,
\end{equation}
which is characterized by only one of the quantum numbers $p_\mu$. We can now introduce a flux via twisted boundary conditions as
\begin{align}
&\ket{\tilde{\Phi}^R(\theta)} = \sum_\mu s_\mu e^{-i\theta \hat{N}^{A_L}} \ket{\psi_\mu}_A \otimes \ket{\psi_\mu}_B. \nonumber\\
&\ket{\tilde{\Phi}^L(\theta)} = \sum_\mu l_\mu e^{-i\theta \hat{N}^{A_L}} \ket{\phi_\mu}_A \otimes \ket{\phi_\mu}_B,
\end{align}
Consider now the biorthogonal reduced density matrix and its eigenstates,
\begin{equation}
\rho_A = \sum_\mu \rho_\mu \ket{\rho_\mu^R}\bra{\rho_\mu^L}
\end{equation}
The right and left eigenstates are not orthogonal but they are linearly independent and therefore form a basis. We can express the Schmidt states in A in terms of the eigenstates of the reduced density matrix and obtain
\begin{align}
&\ket{\tilde{\Phi}^R(\theta)} = \sum_\mu s_\mu e^{-i\theta \hat{N}^{A_L}} (\sum_\nu \psi^\nu_\mu \ket{\rho_\nu^R}) \otimes \ket{\psi_\mu}_B. \nonumber\\
&\ket{\tilde{\Phi}^L(\theta)} = \sum_\mu l_\mu e^{-i\theta \hat{N}^{A_L}} (\sum_\nu \phi^\nu_\mu \ket{\rho_\nu^L}) \otimes \ket{\phi_\mu}_B,
\end{align}

Lets consider again the biorthogonal reduced density matrix, defined as
\begin{align}
\rho_A =& \Tr_B\left[\ket{\tilde{\Phi}^R(\theta)}\bra{\tilde{\Phi}^L(\theta)}\right] \nonumber\\
=& (\sum_{\mu\nu} s_\mu e^{-i\theta N^{A_L}_{\rho_\nu}} \psi^\nu_\mu \ket{\rho_\nu^R} \sum_{\mu'\nu'} l_{\mu'}^* e^{i\theta N^{A_L}_{\rho_{\nu'}}} \phi^{\nu' \, \ast}_{\mu'} \bra{\rho_{\nu'}^L} )\nonumber\\
&\cross \bra{\phi_{\mu'}}_B\ket{\psi_\mu}_B\nonumber \\
=& \sum_{\nu \nu'}\left(\sum_{\mu \mu'} s_\mu l^*_{\mu'} \psi^\nu_\mu \phi^{\nu' \, \ast}_{\mu'}\bra{\phi_{\mu'}}_B\ket{\psi_\mu}_B\right)\nonumber\\
& \cross  e^{-i\theta N^{A_L}_{\rho_{\nu}}} e^{i\theta N^{A_L}_{\rho_{\nu'}}} \ket{\rho_\nu^R}\bra{\rho_{\nu'}^L}
\end{align}
and therefore we can identify the eigenvalues of the reduced density matrix as
\begin{equation}
\left(\sum_{\mu \mu'} s_\mu l^*_{\mu'} \psi^\nu_\mu \phi^{\nu' \, \ast}_{\mu'}\bra{\phi_{\mu'}}_B\ket{\psi_\mu}_B\right) = \rho_\nu \delta_{\nu \nu'}.
\end{equation}
Note that we assume $\ket{\rho_\nu^L}$ and $\ket{\rho_\nu^R}$ have the same number of electrons in $A_L$. This is trivial given that they are eigenstates of $\hat{N}^{A_L}$ and the eigenvalues are real. 
As pointed out previously \cite{herviou_entanglement_2019}, in non-Hermitian systems there is not a simple relation between the Schmidt coefficients and the eigenvalues of $\rho_A$, nonetheless this relation allows us to still connect the reduced density matrix to the polarization. Note that the reduced density matrix, both eigenvalues and eigenvectors, are independent of the flux. 

We can now proceed to compute the biorthogonal polarization, defined as
\begin{align}
\tilde{P} =& i\oint \frac{d\theta}{2\pi}\bra{\tilde{\Phi}^L(\theta)} \partial_\theta \ket{\tilde{\Phi}^R(\theta)} \nonumber\\
=&i\oint \frac{d\theta}{2\pi} \sum_{\mu'\nu'} l_{\mu'}^* e^{i\theta N^{A_L}_{\rho_{\nu'}}} \phi^{\nu' \, \ast}_{\mu'} \bra{\rho_{\nu'}^L} \nonumber\\
&\cross \sum_{\mu\nu} s_\mu (-i N^{A_L}_{\rho_\nu}) e^{-i\theta N^{A_L}_{\rho_\nu}} \psi^\nu_\mu \ket{\rho_\nu^R}\nonumber\\
&\cross  \bra{\phi_{\mu'}}_B\ket{\psi_\mu}_B \nonumber\\
=&\oint \frac{d\theta}{2\pi}\sum_\nu  N^{A_L}_{\rho_\nu}  \left(\sum_{\mu\mu'}s_\mu \psi^\nu_\mu l_{\mu'}^* \phi^{\nu' \, \ast}_{\mu'}  \bra{\phi_{\mu'}}_B\ket{\psi_\mu}_B \right)\nonumber\\
=&\sum_\nu N^{A_L}_{\rho_\nu} \rho_\nu 
\end{align}

This relation generalizes that obtained in Reference \cite{Zelatel_flux_2014} to non-Hermitian systems. In a previous work we showed that, starting from the last line in the equation above and using Peschel trick, for Hermitian systems one can express the polarization as
\begin{align}\label{eq:pol_eos}
    \tilde{P} = \sum_{\mu \in A_L} \xi_\mu,
\end{align}
where $\xi_\mu$ are the eigenvalues of the subsystem correlation matrix, $C_A$, and the sum goes through all the eigenstates that localize near the seam. Since it has been shown that Peschel trick still works for non-Hermitian systems, Eq.\eqref{eq:pol_eos} is also valid in this case.

The assumption that the two virtual cuts are independent of each other is however not valid for point-gapped systems. The reason can be seen in Fig.~\ref{fig:point_gapped_transition}(b), where all eigenstates of the subsystem correlation matrix have a bulk contribution, as opposed to the Hermitian or line-gapped cases.

% - % - % - % - % - % - % - % - % - % - % - % - % - % - % - % - % - % - % - % - % - % - % - % - % - % - % - % - % - % - % - % - % - % - % - % - % - % 

\section{EOS dependence on the ground state}
\label{app:ground_state}

\begin{figure}[t]
\centering
\includegraphics[width=\columnwidth]{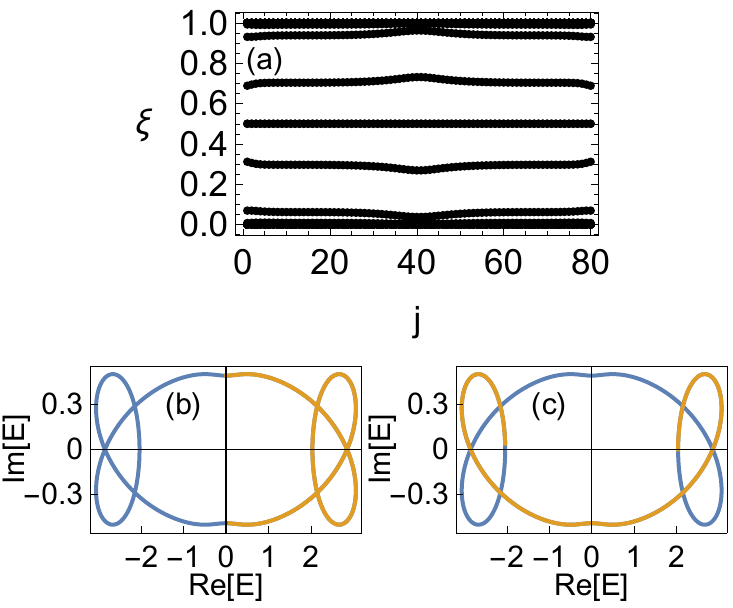}
\caption{(a) EOS for the topological point-gapped phase at $t'=-2t,m=t,\gamma=t/2$ and $\kappa=0$ for different ground states where states are occupied in a connected region characterized by $j$. (b) and (c) energy spectrum for this phase. In yellow we show the occupied states, for $j=0$ and $j=L/2$, respectively.}
\label{fig:point_rotate}
\end{figure}

There is an arbitrariness in how the ground state is selected for non-Hermitian systems.
This is especially problematic for point-gapped phases, where the two bands combine into one. In section~\ref{sec:PG} we showed the results obtained for a half-filled state obtained by occupying states with $\Re[E]<0$. In Fig.~\ref{fig:point_rotate}(a) we show how the EOS (obtained for parameters $t'=-2t,m=t,\gamma=t/2$ and $\kappa=0$) remains qualitatively the same as long as we choose a ground state that is obtained by filling single-particle states in a connected region.
Two such choices are depicted in Fig.~\ref{fig:point_rotate}(b) and (c), where the occupied states are shown in yellow. 
We can define a family of ground states by parametrizing the band by $j\in[0,2L]$, and define a ground state for each starting point of the occupied (yellow) region (note, the ground state in (b) corresponds to $j=0$, while that in (c) to $j=L/2$).

We plot the corresponding EOS as a function of starting point $j$ in Fig.~\ref{fig:point_rotate}(a). As we can see, the EOS remains qualitatively the same. 
Most importantly, the topological state appears at $1/2$ regardless of how we choose the ground state. There is a small discrepancy between different ground states (e.g. between $j=0$ and $j=L/2$), which is due to a finite-size effect.
The dependence on entanglement features was also studied in Ref.~\cite{Guo_entanglement_2021}, where they focused on the entanglement entropy of two different ground states -- filling all states with $\Re[E]<0$ vs. those with $\Im[E]<0$. There, the authors found some difference in the entropy of both states. This is in perfect agreement with our results, as the entropy is an extensive quantity and even $1/N$ effects in the EOS will contribute. 

\end{document}